\documentclass[aps,prl,preprint,groupedaddress]{revtex4}

\usepackage{graphicx}  
\makeatletter
\usepackage[sort&compress]{natbib}
\begin{document}

\title{Location-Dependent Communications using Quantum Entanglement}

\author{Robert~A.~Malaney, School of Electrical Engineering and Telecommunications, University of New South Wales,  NSW 2052, Australia.
r.malaney@unsw.edu.au}

\begin{abstract}

 The ability to unconditionally verify the location of a
communication receiver would lead to a wide range of new security
paradigms. However, it is known that unconditional location
verification in classical communication systems is impossible. In
this work we show how unconditional location verification can be
achieved with the use of quantum communication channels. Our
verification remains unconditional irrespective of the number of
receivers, computational capacity, or any other physical resource held by an adversary. Quantum
location verification represents a new application of quantum
entanglement that delivers a feat not possible  in the
classical-only channel. For the first time, we possess the ability
to deliver real-time communications viable only at specified
geographical coordinates.

\end{abstract}
\maketitle

 The ability to offer a real-time communication channel whose viability is unconditionally a function of the receiver location would offer a range of new information security paradigms and applications. In particular, there are a range of industries and organizations that clearly would be interested in delivering information content in the sure knowledge a recipient receiver is at an \emph{a-priori} agreed upon location (\emph{e.g.} see discussions in \cite{denning,classic1,malaney,classic}). The ability to guarantee location-sensitive communications requires unconditional (independent of the physical resources held by an adversary) location verification. However, in the classical-only channel such unconditional location verification is impossible.  The finite speed of light  can only be used to bound the minimum (but not the maximum) range a receiver is from some reference station.
 Add to the mix that classical information can be copied, and that an adversary can possess unlimited receivers (each of which can be presumed to possess unlimited computational capacity), it is straightforward to see why classical-only unconditional location verification is impossible. It is the purpose of this work to show how the introduction of quantum entanglement into the communication channel overcomes the above concerns, providing for the first time an unconditional location verification protocol.

 Quantum
teleportation  \cite{tele1},  the transfer of unknown quantum
state information, is now experimentally verified through a host
of
 experiments,  \emph{e.g.} \cite{tele2,tele5}. In addition,
the key resource underpinning teleportation, quantum entanglement,
has been experimentally verified over very large ranges. An
entanglement measurement over  $144$km, achieved recently using
optical free-space communications between two telescopes
 \cite{tenrife}, proves the validity of ground-station to satellite
quantum communications, and is widely seen as a major step in the
path towards a global quantum communications network. In such a
network it is envisaged that  a combination of satellite and fiber
optic links  will interconnect a multitude of quantum nodes,
quantum devices   and quantum computers. In optical fiber,
transmission of entangled photons is limited to about 100km by
losses and de-coherence effects, \emph{e.g.} \cite{fibre4}.
Communications over fiber  beyond this range will make use of
either quantum repeaters  \cite{repeaters}, or the trusted relay
paradigm used in a recent deployment of an eight-node quantum
network  \cite{trusted}.

Experimental verification of quantum superdense coding
\cite{dense1} has also been achieved through a series of
experiments, \emph{e.g.}  \cite{dense2,dense6}. In superdense
coding, two bits of classical information can be transferred at
the cost of only one qubit.

Teleportation and superdense coding are strongly related, and
indeed they are often considered as protocols which are the
inverse of each other, differing only in how and when they utilize
quantum entanglement. Quantum location verification can be
considered a new protocol that differs again in how and when it
uses quantum entanglement.

The principal condition  for  unconditional location
verification is that;
only a device at \emph{one} unique location (the authorized location) is
able to, \emph{immediately} and \emph{correctly},  respond
to signals received from multiple reference stations. In the classical-only channel this condition can never be unconditionally guaranteed.
However, as we now show,
 with the introduction of  quantum communication
channels the  condition necessary for unconditional location
verification can in fact be guaranteed.

Consider some  reference stations at publicly known locations, and
a device which is not a reference station (Cliff) that is to be verified at a publicly known
location $(x_v,y_v)$. Let us assume that processing times, such as those due to local quantum measurements, are negligible (we discuss later the minor impact of this). We
also assume that the reference stations are authenticated and
share secure communication channels between each other via quantum
key distribution (QKD) \cite{qcd1,qcd2},  and that
 all  classical communication between Cliff and the reference
 stations occurs via wireless channels. The use of wireless communications is important
 since we will require  the time delay of all classical
 communications to be set by the line-of sight-distance between
 transceivers divided by $c$ (light speed in
 vacuum).

For two dimensional location verification we require a minimum of
three reference stations. Consider $N$ maximally entangled
multipartite systems available to a network possessing $k$
reference stations. Consider also that each of the
 multipartite systems comprises $k$ qubits,
  with each reference station initially holding one qubit from each of the $N$ systems.
  The $2^k$  orthogonal basis states of each multipartite state can be written
  \begin{equation}
\left| S_b \right\rangle  = \frac{1}{{\sqrt 2 }}\left( {\left| a
\right\rangle _1  \otimes \left| a \right\rangle _2  \ldots \left|
a \right\rangle _k  \pm \left| a \right\rangle _1  \otimes \left|
a \right\rangle _2  \ldots \left| a \right\rangle _k } \right)
 \label{eq:1a}
\end{equation}
where  ${b = 1,...2^k }$, and the states $ \left| a \right\rangle$
 represent $\left| 0 \right\rangle$ or $\left| 1
\right\rangle$ with the index on the state labelling the location (ignoring any \emph{null} state).

Transformation between the basis states can be achieved by a set
of $2^k$ unitary transformations  induced on the locally held
qubits. By this means a $k$-bit message, per entangled state, can
be transferred from the stations to Cliff.
 This is achieved using superdense coding in which the stations
 encode each message to a specific basis state $ \left|
S_b \right\rangle $, with Cliff decoding the message via a quantum
measurement that deterministically  discriminates all possible
basis states (the state  $ \left|
S_b \right\rangle $ is sent directly to Cliff via quantum channels connected to the reference stations).

Quantum location verification builds on this concept of state
encoding with one key addition. It must be the case that
deterministic discrimination amongst the encoded states is
possible, \emph{within a pre-described time bound at only one
location}. This can be achieved if the  $2^k$ states which encode
the $k$-bit messages are made \emph{non-orthogonal} by  the
introduction of an additional local unitary transformation at each
reference station. Let these additional transformations be
labelled $U_i^{r}$, where ${r = 1,...k }$ indexes the reference
station, and $i=1...N$ references the specific multipartite state
to which the local transformation is applied.

 Consider the $ith$ encoded multipartite state in which
a
 $k$-bit message is encoded as $\left| {S_b
}\right\rangle$. Then on application of the additional
transformations a new state $ \left| {\Upsilon _i } \right\rangle
= U_i^1 \otimes U_i^2 \otimes ...U_i^k \left| {S_b }
\right\rangle$ is produced. Our requirement is that $ \left\langle
{{\Upsilon _i }}
 \mathrel{\left | {\vphantom {{\Upsilon _i } {\Upsilon _j }}}
 \right. \kern-\nulldelimiterspace}
 {{\Upsilon _j }} \right\rangle  \ne 0$ when
 $
\left| {\Upsilon _i } \right\rangle  \ne \left| {\Upsilon _j }
\right\rangle$.
 Ideally, the unitary matrices $U_i^r$ are chosen
so that upon measurement  of $ \left| {\Upsilon _i } \right\rangle
$ in a  measurement basis $ \left| {S_1 } \right\rangle ,\left|
{S_2 } \right\rangle  \ldots \left| {S_{2^k } } \right\rangle $,
the probability of collapse to each  basis state is approximately
equal ($1/2^k$).

For quantum location verification to be unconditional it must be
impossible for an adversary to map the values of $U_i^{r}$ to
specific $k$-bit messages (in our protocol all matrices $U_i^{r}$
and all $k$-bit messages are ultimately sent over a classical
channel). This means that there must be some form of randomness
applied to the selection of each $U_i^{r}$.
 One strategy that
provides for both a random selection mechanism,
 and  the required
non-orthogonal behavior between the states $ \left| {\Upsilon _i }
\right\rangle $, is  to allow the $U_i^{r}$ to be constructed from
four random real parameters ($\alpha,\beta,\gamma, \phi)$. The
unitary matrix at each reference station can then be implemented
as
\begin{equation}
U = e^{i\phi } R_z \left( \alpha \right)R_y \left( \beta
\right)R_z \left( \gamma  \right) , \label{eq:1r}
\end{equation}
where the rotations $R$ are given by
\[
R_y (\theta ) = e^{ - i\theta \sigma _y /2} {\rm{ \  and \
}}R_z (\theta ) = e^{ - i\theta \sigma _z /2} ,
\]
and with the $\sigma$'s representing the Pauli operators.
Classical communication of the additional matrices can be achieved in many ways, such as passing of experimental instructions (e.g. duration of laser pulses), indexing of a large number of matrices,
or as  a transfer of matrix element information. The latter, which we adopt here,  involves
the transmission of the values ($\alpha,\beta,\gamma, \phi)$
adopted for each $U_i^r$. Although finite bandwidth of the classical channel limits the precision of this information transfer -  required precision is available at the cost of additional bandwidth. In actual deployment, any global phase
can be ignored. We discuss later a pragmatic implementation
strategy leading to an outcome effectively the same as the outcome
derived from Eq.~(\ref{eq:1r}).

 The location verification proceeds by the encoding of a secret
sequence onto a set of $N$ entangled systems $ \left| {\Upsilon
_{i=1...N} } \right\rangle $, transmission of each $ \left|
{\Upsilon _{i} } \right\rangle $ to Cliff via quantum channels,
followed by transmission of the unitary matrices $U_i^{r}$
(\emph{i.e.} the set ($\alpha,\beta,\gamma, \phi)$)  to Cliff by
classical channels. Upon receiving this quantum and classical
information Cliff can decode and broadcast the decoded sequence
via the classical channel. Given that information transfer over
the classical channel proceeds at a velocity $c$, location
information becomes unconditionally verifiable (as explained
later). Ultimately, the verification is based on the inability to
clone deterministically the set $ \left| {\Upsilon _i }
\right\rangle $ with fidelity one. Although cloning with lower
fidelities is possible, confidence levels on the location
verification can be increased to any arbitrary level by increasing
$N$.

We now outline the protocol in more detail using well known
maximally entangled states. For clarity, we proceed with a one
dimensional location verification using just two reference
stations, which we henceforth refer to as Alice and Bob. A
geometrical constraint for one-dimensional location verification
is that the device to be located must lie between Alice and Bob.
That is, $ \tau _{AC} + \tau _{BC} = \tau _{AB} $, where $ \tau
_{AC}$ ($ \tau _{BC}$) is the light travel time between Alice
(Bob) and Cliff, and where $\tau_{AB}$ is the light travel time
between Alice and Bob.

 Let Alice share with
Bob a set of \emph{N} maximally entangled qubit  pairs $ \left|
{\Omega _i^{AB} } \right\rangle$, where the subscript $i = 1
\ldots N$ labels the entangled pairs.
 Let each of the pairs be described by
one of the Bell states $ \left| {\Phi ^ \pm  } \right\rangle =
\frac{1}{{\sqrt 2 }}\left( {\left| {00} \right\rangle  \pm \left|
{11} \right\rangle } \right)$, $ \left| {\Psi ^ \pm  }
\right\rangle  = \frac{1}{{\sqrt 2 }}\left( {\left| {01}
\right\rangle  \pm \left| {10} \right\rangle } \right)$, with the
first  qubit being held by Alice  and the second by Bob.
 We will
assume an encoding  ($ 00 \to \Phi ^ +$ \emph{etc.}) that is
public.

 Without loss of generality we can
assume all pairs are initially in the state $ \left| {\Phi ^ + }
\right\rangle$. After the encoding of a sequence onto a series of
entangled pairs, Alice and Bob apply an additional random unitary
transformation $U^A_i$ and $U^B_i$, respectively, to their local
qubit from each pair.
 As a consequence, the entangled pairs held
by Alice and Bob now form a non-orthogonal set,
\begin{equation}
\left| {\Upsilon _i^{AB} } \right\rangle  = U_i^A  \otimes U_i^B
\left| {\Omega _i^{AB} } \right\rangle  \ \ . \label{eq:1}
\end{equation}
For example, for $\left| {\Phi ^ +  } \right\rangle $
Eq.~(\ref{eq:1}) leads to a state $ \frac{1}{{\sqrt 2 }}\left(
{U_i^A \left| 0 \right\rangle _A \otimes U_i^B \left| 0
\right\rangle _B  + U_i^A \left| 1 \right\rangle _A  \otimes U_i^B
\left| 1 \right\rangle _B } \right)$.

\begin{figure}
 \includegraphics[width=3.5in,height=3.4in,clip,keepaspectratio]{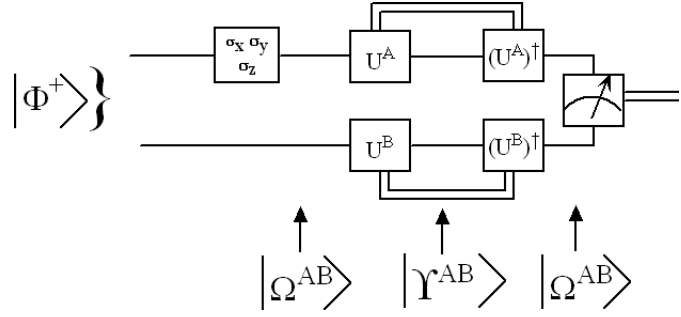}\\
  \caption{Quantum Circuit for Location Verification}
  \label{fig1}
\end{figure}

\noindent A step-by-step exposition of the protocol follows.

\noindent $\bullet$ Step 1: Via a secure channel Alice and Bob
agree on a mutual random bit sequence $S_{ab}$ that is to be
encoded. The encoding is achieved via superdense coding in which
two classical bits are encoded using local unitary operators as
described by $ I\left| {\Phi ^ + } \right\rangle = \left| {\Phi ^
+  } \right\rangle $, $ \sigma _x \left| {\Phi ^ + } \right\rangle
= \left| {\Psi ^ + } \right\rangle$, $ i\sigma _y \left| {\Phi ^ +
} \right\rangle = \left| {\Psi ^ -  } \right\rangle$, and $ \sigma
_z \left| {\Phi ^ +  } \right\rangle = \left| {\Phi ^ -  }
\right\rangle$.  For each pair of entangled qubits, Alice and Bob
also agree who will induce the necessary unitary operation on
their local qubit in order to encode sequential two-bit segments
of $S_{ab}$.

\noindent $\bullet$ Step 2: Prior to the transmission of  any
qubit the transformation $ \left| {\Omega _i^{AB} } \right\rangle
\to \left| {\Upsilon _i^{AB} } \right\rangle$ as described by
Eq.~(\ref{eq:1}) is induced. This set is then transmitted by Alice
and Bob to Cliff via two separate quantum channels.

\noindent $\bullet$ Step 3: Alice  and Bob communicate to Cliff,
via separate classical channels, the random matrices $U^A_i$ and
$U^B_i$ used to form the set $ \left| {\Upsilon _i^{AB} }
\right\rangle $. This
 classical information is transmitted in a synchronized manner to Cliff
such that for each value of $i$ the $U^A_i$ sent by Alice and  the
$U^B_i$ sent by Bob arrive simultaneously at Cliff's publicly
announced location $(x_v,y_v)$. It is also ensured that this
classical information is received at Cliff \emph{after} the
arrival of the corresponding qubit pair of $ \left| {\Upsilon
_i^{AB} } \right\rangle$.

\noindent $\bullet$ Step 4: Upon receipt of each matrix pair
$U^A_i$, $U^B_i$, Cliff undertakes the transform $\left( {U_i^A
\otimes U_i^B } \right)^\dag  \left| {\Upsilon _i^{AB} }
\right\rangle \to  \left| {\Omega _i^{AB} } \right\rangle $
 before taking a Bell State Measurement
(BSM) in order to determine the two-bit segment encoded in the
entangled pair. Cliff then  immediately broadcasts (classically)
the decoded two-bit segment back to Alice and Bob.

\noindent $\bullet$ Step 5: Alice checks that the sequence
returned to her by Cliff is correctly decoded and notes the
round-trip time for the process. Likewise Bob. Alice and Bob can
then compare their round-trip times to Cliff (2$\tau_{AC}$ and
2$\tau_{BC}$) in order to verify consistency with Cliff's publicly
reported location $(x_v,y_v)$.

\noindent The quantum circuit for the one dimensional quantum
location verification just described is given in fig.~1.

Quantum location verification is independent of the physical
resources an adversary may possess.  In the classical-only channel an
adversary can place co-operating devices closer to reference
stations and then delay responses in order to defeat any location
verification (\emph{e.g.} \cite{classic1,classic}). Attempts to remedy this problem by making devices unique and tamper-proof are clearly limited (see discussion in \cite{classic1}), and cannot provide for unconditional security.

 However, in quantum
verification multiple devices are of no value. In order to decode
immediately, Cliff's receiver must possess \emph{all} the qubits
that comprise each  entangled state. Cliff cannot distribute
copies of his local qubits to other devices due to the no-cloning
theorem \cite{noclon}. The key point is that
 for any given location
$(x_v,y_v)$ that is to undergo a verification process, one can
always find placements for the reference stations such that no
other location can be \emph{simultaneously} closer to all of the
reference stations than $(x_v,y_v)$ (\emph{e.g.} recall the geometrical constraint in our one dimensional verification).
 This being the case, an
adversary with no device at the location being verified cannot
pass the verification test. Even if the adversary possess multiple
receivers, an additional round-trip communication time between his
devices will be required for decoding. This will result in a
 round-trip time between at least one reference station
and the location $(x_v,y_v)$  being larger than expected. In
classical verification the \emph{round-trip} communication between
the  adversary's devices is not required.

 Extension of the
one-dimensional location verification protocol  to two-dimensional
verification could be a straightforward application of additional
bipartite entanglement between Alice and some third reference
station, say Dan. This can be achieved by introduction of a new
set of Bell states shared between Alice and Dan, with the protocol
following a similar exposition to that given. However, perhaps a
more elegant solution  is the use of multipartite entangled
states. For example, consider a Green-Horne-Zeilinger (GHZ)
\cite{ghz1,ghz2} state in which three qubits are maximally
entangled, such as $ \left| S \right\rangle ^ +   =
\frac{1}{{\sqrt 2 }}\left( {\left| {000} \right\rangle  + \left|
{111} \right\rangle } \right)$. Transformation from this GHZ basis
state to one of the eight  basis states is achieved by the set
of transforms    $U_{GHZ}=(\sigma _z \otimes \sigma _z ,I_2
\otimes \sigma _z ,i\sigma _y \otimes \sigma _z ,\sigma _x \otimes
\sigma _z ,I_2 \otimes \sigma _x ,\sigma _z \otimes \sigma _x ,
\sigma _x \otimes \sigma _x ,i\sigma _y \otimes \sigma _x $),
where the first (second) operator acts on the first (second) qubit
\cite{ghz6}. A step-by-step quantum location verification using
such tripartite states proceeds in similar manner to the bipartite
protocol.

Clearly, a security threat to the protocol is the potential
ability of an adversary who is in possession of an optimal cloning
machine, redistributing the set $\left| {\Upsilon _i }
\right\rangle$ to other devices. If cloning were exact,  the
verification test would fail because the \emph{round-trip}
communication between the devices (needed to decode) would not be
required.
 However, optimal cloning of the set $\left| {\Upsilon
_i } \right\rangle$ can be described by the fidelity, $F_c$,
between this set and a cloned set. This is known to be upper
bounded by $F_c
 \approx 0.7$ for bipartite entanglement and $F_c
 \approx 0.6$ for tripartite entanglement  \cite{clone2,clone3}.
 As such, for  a series of two-bit messages encoded in $N=100$ bipartite states, an optimal cloning machine would have a probability of 1 in
$10^{16}$ of passing
 the verification system even though not at the authorized location. For 100 three-bit messages encoded in
 tripartite states this decreases to a probability of  1 in $10^{22}$.
 Arbitrary smaller probabilities are achieved exponentially in $N$.

A key aspect of our protocol is rapid implementation of the random
unitary matrices, $U_i^{r}$, at the reference stations.  One
pragmatic strategy that provides for both a random selection
mechanism,
 and  the required
non-orthogonal behavior between the states $\left| {\Upsilon _i }
\right\rangle$, is  to allow the $U_i^{r}$ to be constructed from
 random permutations of the Hadamard gate $H$, and the $\pi /8$
gate $T$. It is known that any single-qubit unitary operation can
be approximated to arbitrary accuracy from $H$ and $T$ gates
(\emph{e.g.} \cite{chang}), and that standard optical devices can
be deployed to induce such gates on polarized photons.
In simulations we have explored permutations of the $T$ and $H$
gates as a means of producing the random transforms needed to
remove the orthogonality of the original basis. A series of random
permutations leading to gates of the form $TTHTHHTTH....$ were
performed, and the average orthogonality of the set $\left|
{\Upsilon _i } \right\rangle$ measured. It was found that even
with gates using only 5 random combinations (e.g. $THHTH$) the
required non-orthogonal properties between the states $\left|
{\Upsilon _i } \right\rangle$ was achieved -
 with the average fidelity
between any two states being $F\sim 0.3$. Similar fidelities were
found using the random matrix formulation of Eq.~(\ref{eq:1r}).

The new quantum protocol we have outlined is aimed at networks in
which the quantum channel utilizes fiber and the classical channel
utilizes wireless communications. The protocol requires the  use
of random transformations at the reference stations, and the
presence of efficient millisecond quantum memory at the receiver
(see \cite{laur} for state-of-the-art implementation of quantum
memory at telecommunication wavelengths). However, implementation
of our protocol is simpler  when it is assumed that qubits in the
quantum channel move with velocity $c$, as  no additional
transformations are required, and the need for quantum memory is
negated (in many set-ups). In such a circumstance the
one-dimensional verification protocol would follow a set-up
similar to that utilized in
 recent experiments on
entanglement swapping \cite{swap}. In \cite{swap}, a BSM via
linear optics is conducted on a series of entangled photons
arriving from different synchronized pulsed sources. Coincidence
counting is achieved within the nanosecond range. Using similar
techniques, an implementation of location verification over tens
of km, to an accuracy of meters is currently possible. Any
relaxation of our initial assumptions such as zero processing
time, will manifest itself in a (determinable) reduction in the
accuracy of the location being verified. Note that even though we
have described our protocol under the
 assumption that all four Bell states can be discriminated in the
 BSM - this is not a requirement. When using linear optics for BSM
  only two Bells states can be discriminated (deterministically). In
 this case our encoding scheme would need to be adjusted to a three
 message encoding. This has the minor effect of a drop in the channel
 capacity.

Clearly there are many variants on our protocol, such as the use
of teleportation, the use of other entanglement degrees of
freedom, and the use of entanglement swapping between
 the reference stations and the device.  For example, a modified
verification protocol that uses entanglement swapping can be
constructed that entirely negates the requirement for direct
transfer of qubits between the reference stations and the device.
Location verification would then be possible in a
satellite-to-device communications system, provided the satellite
and the device shared \emph{a-priori} an entangled resource stored
in quantum memory.

 Quantum location verification could greatly assist in the
authentication of devices within large-scale multihop quantum
networks \cite{longline}. Current quantum authentication
techniques require the distribution of secret keys distributed
\emph{a priori} amongst potential users \cite{whyqcd}. However,
such keys, whether classical bits or entangled qubits, are subject
to unauthorized re-distribution. We also note that quantum
location verification can be used within other data-delivery
protocols in which real-time data transfer can be communicated to
a device successfully \emph{only if} that device is at a specific
location. The location verification can be monitored continuously
in real time, halting any real-time data transfer upon violation
of the verification procedures. An adversary could not continue to
receive real-time data without one of his devices being at the
specified location.

 Quantum
location verification represents a new application in the emerging
field of quantum communications. It delivers an outcome not
possible in the classical-only channel. For the first time, we
possess the ability to unconditionally authenticate a
communication channel based on the geographical coordinates of a
receiver.

\section*{Acknowledgment}
This work has been supported by the University of New South Wales.

\end{document}